\documentclass[journal]{IEEEtran}
\usepackage{xcolor} 

\usepackage{generic}
\usepackage{amsmath,amssymb,amsfonts}
\usepackage{algorithm2e}
\usepackage{graphicx}
\usepackage{multirow}
\usepackage{textcomp}

\def\BibTeX{{\rm B\kern-.05em{\sc i\kern-.025em b}\kern-.08em
    T\kern-.1667em\lower.7ex\hbox{E}\kern-.125emX}}
\usepackage[style=ieee]{biblatex}
\begin{document}
\title{Dynamic Topic Analysis in Academic Journals using Convex Non-negative Matrix Factorization Method}
\author{Yang Yang, Tong Zhang, Jian Wu, and Lijie Su
% \thanks{This research was financially supported by the Major Program of National Natural Science Foundation of China under Grant 72192830 and Grant 7219283l, the National Natural Science Foundation of China under Grant 72072029, and the 111 Project under Grant B16009. Paper no. xxx (Corresponding author: Jian Wu)}
\thanks{Yang Yang was with the National Frontiers Science Center for Industrial Intelligence and Systems Optimization, Northeastern University, Shenyang, China. He is now with China Criminal Police University, Shenyang, China (e-mail: yangyang@cipuc.edu.cn). }
\thanks{Tong Zhang was with the Center for Advanced Process Decision-making, Carnegie Mellon University, Pittsburgh, PA, USA. He is 
now with the Beijing Institute for General Artificial Intelligence, Beijing, China (e-mail: tongzhang1990@gmail.com).}
\thanks{Jian Wu is with the Liaoning Key Laboratory of Manufacturing System and Logistics Optimization, Northeastern University, Shenyang 110819, China, Shenyang, China (e-mail: wujian@mail.neu.edu.cn).}
\thanks{Lijie Su is with the Liaoning Engineering Laboratory of Data Analytics and Optimization for Smart Industry, Northeastern University, Shenyang, China (e-mail: sulijie@ise.neu.edu.cn).}
}

\maketitle

\begin{abstract}
% With the exponential growth of text data, the need for efficient and dynamic methods to extract valuable insights has become increasingly critical. Dynamic topic analysis offers a robust methodology to capture and understand the temporal evolution of topics in large-scale datasets. This work proposes a two-stage dynamic topic analysis framework that integrates convex optimization to enhance topic consistency, sparsity, and interpretability. In Stage 1, a two-layer non-negative matrix factorization (NMF) model is employed to extract annual topics and identify relevant keywords. In Stage 2, a convex optimization algorithm is explored for dynamic topic refinement based on the convex NMF model, further enhancing and integrating the identified topics. Through a dynamic analysis of IEEE journal abstracts from 2004 to 2022, the proposed method effectively identifies and quantifies emerging topics such as COVID-19 and digital twins. By optimizing the sparsity differences in the clustering feature space between traditional and emerging research topics, this approach provides deeper insights into topic evolution and ranking analysis. Furthermore, the NMF-cNMF model demonstrates higher stability in topic consistency. 
% At sparsity levels of 0.4, 0.6, and 0.9, the proposed approach enhances the stability of topic rankings improves by 24.51\%, 56.60\%, and 36.93\%, respectively. The source code is available at https://github.com/meetyangyang/CDNMF.

With the rapid advancement of large language models, academic topic identification and topic evolution analysis are crucial for enhancing AI's understanding capabilities. Dynamic topic analysis provides a powerful approach to capturing and understanding the temporal evolution of topics in large-scale datasets. This paper presents a two-stage dynamic topic analysis framework that incorporates convex optimization to improve topic consistency, sparsity, and interpretability. In Stage 1, a two-layer non-negative matrix factorization (NMF) model is employed to extract annual topics and identify key terms. In Stage 2, a convex optimization algorithm refines the dynamic topic structure using the convex NMF (cNMF) model, further enhancing topic integration and stability. Applying the proposed method to IEEE journal abstracts from 2004 to 2022 effectively identifies and quantifies emerging research topics, such as COVID-19 and digital twins. By optimizing sparsity differences in the clustering feature space between traditional and emerging research topics, the framework provides deeper insights into topic evolution and ranking analysis. Moreover, the NMF-cNMF model demonstrates superior stability in topic consistency. At sparsity levels of 0.4, 0.6, and 0.9, the proposed approach improves topic ranking stability by 24.51\%, 56.60\%, and 36.93\%, respectively. The source code is available at https://github.com/meetyangyang/CDNMF.
\end{abstract}

\section{Introduction}
\begin{figure}[!t]
\centering
\includegraphics[width=\linewidth]{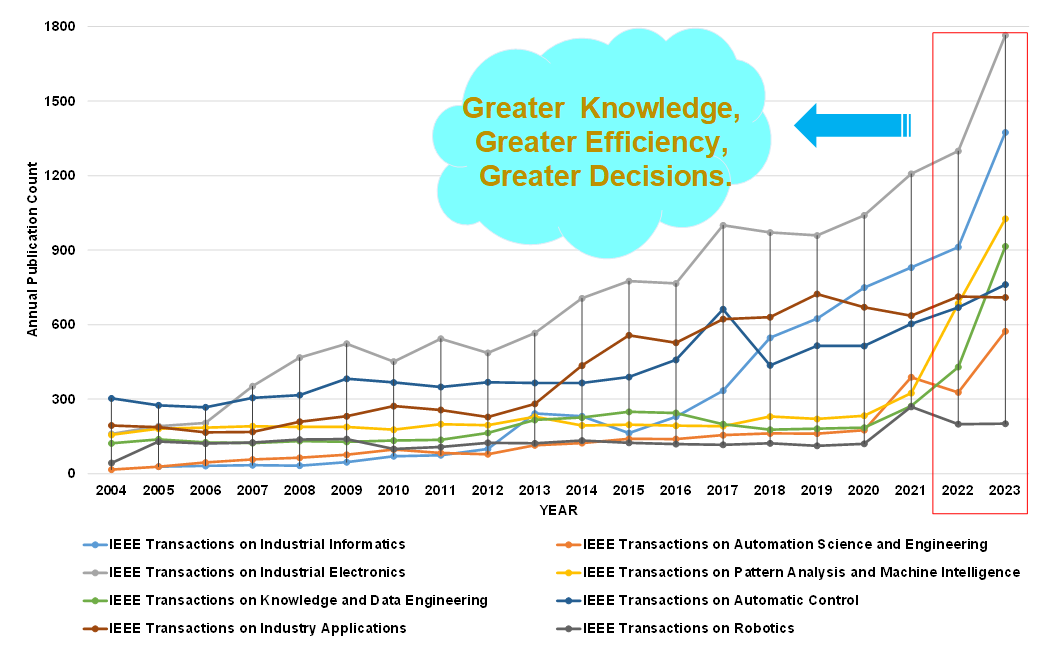}
\caption{Number of papers published in IEEE journals}
\label{fig:background}
\end{figure}

\IEEEPARstart{T}{he} exponential growth of academic publications presents significant challenges for researchers attempting to identify emerging patterns and trends efficiently. As illustrated in Fig. \ref{fig:background}, the increasing volume of IEEE journals highlights the limitations of conventional NLP methods in handling such complex literature networks, often leading to cognitive overload and fragmented knowledge discovery \cite{chen2014tag}. While open-source large language models (LLMs) like DeepSeek R1 \cite{guo2025deepseek} have demonstrated remarkable capabilities in general text processing, their effectiveness in scholarly topic analysis remains constrained by three critical gaps: (1) Lack of domain-specific temporal modeling mechanisms to capture knowledge evolution trajectories; (2) Over-reliance on parametric memorization rather than interpretable pattern mining; (3) Inadequate handling of sparse academic discourse through monolithic architectures.

These limitations become particularly evident when examining DeepSeek's Mixture-of-Experts (MoE) paradigm. Although MoE structures excel at task specialization through dynamic routing, their black-box nature conflicts with academic analysis' core requirement for transparent knowledge decomposition. This contrast motivates our adoption of Non-negative Matrix Factorization (NMF) as a methodological anchor – its additive parts-based representation naturally aligns with the compositional characteristics of research topics. By further integrating convex optimization constraints into the factorization process, we establish mathematically traceable relationships between temporal slices while maintaining essential sparsity characteristics.

\begin{figure*}[!t]
\centering
\includegraphics[width=\linewidth]{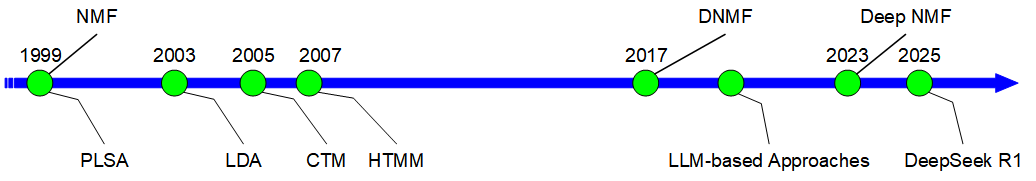}
\caption{Some representative topic modeling methods in dynamic topic analysis}
\label{fig:tm_his}
\end{figure*}
Dynamic topic analysis has become an indispensable tool to address these challenges, enabling researchers to uncover hidden topics and understand their temporal evolution. By leveraging unsupervised models such as topic modeling and matrix factorization techniques, researchers can extract meaningful insights from large-scale datasets. However, challenges remain, including ensuring the stability and sparsity of identified topics while balancing interpretability and model complexity.

Significant progress has been made in dynamic topic modeling, as shown in Fig. \ref{fig:tm_his}. Early generative models such as Latent Dirichlet Allocation (LDA)~\cite{blei2003latent} and Probabilistic Latent Semantic Analysis (PLSA)~\cite{hofmann1999probabilistic} laid the foundation for unsupervised topic discovery. Extensions such as Hidden Topic Markov Models (HTMM)~\cite{wang2006topics} and Correlated Topic Models (CTM)~\cite{lafferty2005correlated} introduced temporal correlations, yet retained inherent limitations in modeling discontinuous paradigm shifts common in academic domains. While achieving contextual awareness, modern large language models(LLM) and LLM-based approaches often produce chronologically inconsistent results due to their static knowledge cutoff dates and lack of explicit temporal modeling layers.

Matrix factorization methods, particularly Non-negative Matrix Factorization (NMF)~\cite{lee1999learning}, have emerged as a cornerstone in topic modeling due to their capacity to generate interpretable and sparse topic representations. Recent advancements, such as dynamic NMF (DNMF)~\cite{greene2017exploring} and deep NMF~\cite{wang2023deep}, have substantially improved efficiency and scalability. However, challenges such as parameter selection, data sparsity, and spatiotemporal integration persist.

To address these challenges, this study proposes a two-stage dynamic topic analysis framework that integrates convex optimization to improve topic consistency, sparsity, and interpretability. In the first stage, a dynamic NMF model extracts annual topics and identifies key terms. The second stage refines these results using a convex-optimized NMF model, enhancing stability and structure. Evaluated on abstracts from IEEE journals, the proposed approach effectively identifies traditional and emerging research trends.

The key contributions of this study include:
\begin{itemize}
    \item A two-stage dynamic topic analysis framework combining temporal topic modeling and convex optimization to improve topic consistency and interpretability.
    \item Development of a convex NMF model that enhances the stability and sparsity of topic matrices.
    \item Validation of the proposed method through experiments on IEEE journal abstracts, showcasing its ability to identify emerging trends in academic research.
\end{itemize}

\begin{figure}[!t]
\centering
\includegraphics[width=\linewidth]{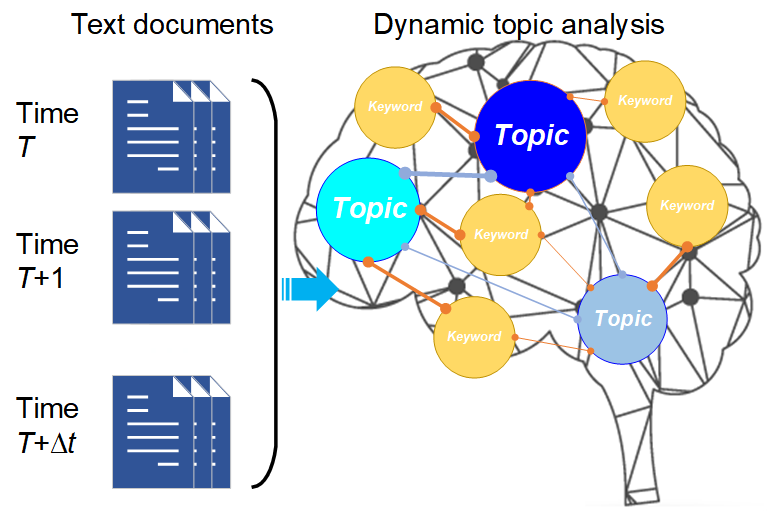}
\caption{An overview of dynamic topic models.}
\label{fig:problem}
\end{figure}

\section{Problem Description}
\subsection{Overview of Dynamic Topic Analysis}

Dynamic topic analysis uncovers the temporal evolution of topics and trends within time-stamped or streaming text data, as illustrated in Fig.~\ref{fig:problem}. Models like dynamic LDA and dynamic NMF are designed to capture temporal topic dynamics and identify emerging trends, offering deeper insights into evolving topics within text corpora~\cite{greene2017exploring,blei2006dynamic,cui2011textflow}. These methods mathematically model the topic generation process, yielding elegant and theoretically grounded representations. However, these models introduce challenges, such as increased computational complexity and susceptibility to overfitting, due to additional parameter estimation~\cite{blei2003latent,lee1999learning}.

Dynamic topic analysis imposes several critical requirements:
\begin{itemize}
    \item \textbf{Sparsity:} The model should generate topic representations with a minimal number of active components to reduce noise and redundancy, thereby improving interpretability and clarity.
    \item \textbf{Stability:} The model must ensure consistent topic distributions across temporal snapshots or parameter configurations, which is crucial for reliable trend analysis.
    \item \textbf{Optimality:} The model should identify topic representations that best explain the underlying text data, facilitating the detection of both dominant and emergent topics.
\end{itemize}

Addressing these requirements is essential for improving the accuracy and usability of dynamic topic models in academic and practical applications.

\subsection{Convex Optimization in Dynamic Topic Analysis}

\begin{figure*}[!t]
\centering
\includegraphics[width=\linewidth]{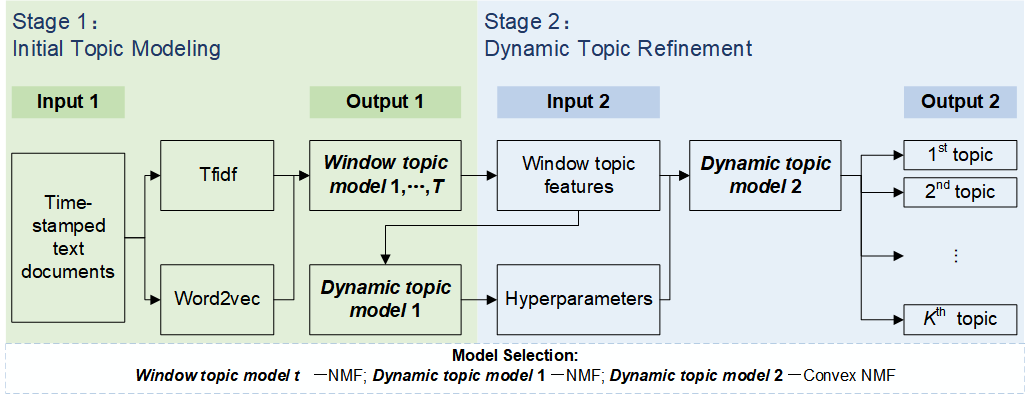}
\caption{Research Framework}
\label{fig:flow}
\end{figure*}

Convex optimization refines dynamic topic models by addressing key challenges, including sparsity, stability, and optimality, thereby enhancing model performance and interpretability:
\begin{itemize}
    \item \textbf{Enhancing Sparsity:}  
    Convex optimization enforces sparsity through non-negativity and sparsity penalties, encouraging topic representations with fewer active components. This reduces noise and enhances interpretability, enabling concise identification of core topics in evolving text corpora.

    \item \textbf{Improving Stability:}  
    Stability ensures consistent topic distributions across temporal snapshots or dataset shifts. Convex optimization achieves this by incorporating regularization techniques that mitigate sensitivity to data and parameter variations. This robustness is vital for accurately tracking the temporal evolution of topics.

    \item \textbf{Achieving Optimality:}  
    Convex optimization guides models to solutions that optimally balance complexity and data fidelity. Techniques like \(L_1\) and \(L_2\) regularization detect subtle yet significant changes in topic prevalence over time, ensuring interpretable and representative results.
\end{itemize}

In conclusion, integrating convex optimization into dynamic topic models addresses traditional limitations, enhancing effectiveness, interpretability, and robustness in topic analysis. Building on these advancements, this study proposes a two-stage methodology based on NMF to analyze academic journals. The details of this methodology are presented in the subsequent sections.

\section{A Two-Stage Dynamic Topic Modeling Approach}

Building on the prior research in academic topic analysis~\cite{zhang_forty_2019}, this study proposes a convex optimization-enhanced two-stage dynamic topic modeling approach. In Stage 1, negative matrix factorization models serve as the foundation for dynamic topic analysis. Stage 2 builds upon the clustering and feature word extraction results derived from the DNMF model in Stage 1.

The proposed approach incorporates convex non-negative matrix factorization (cNMF) to optimize dynamic topic modeling, enhancing its precision and computational efficiency. By leveraging convex optimization techniques, the approach enhances the performance of dynamic topic modeling, improving robustness and adaptability in real-world applications. These enhancements make it particularly well-suited for academic corpora and topic analysis.

\subsection{Stage 1: Initial Topic Modeling}

As illustrated in Fig.~\ref{fig:flow}, the proposed two-stage framework based on the DNMF model yields two primary outputs: (1) window topic models with their corresponding features, and (2) dynamic topic models with optimized hyperparameters.

The topic model generated for each time period is known as the window topic model. These window topic models are dynamically integrated across time windows to construct a comprehensive topic model, known as the dynamic topic model. Typically, the hyperparameters of topic models require fine-tuning to achieve optimal topic analysis results.

\subsubsection{Window Topic Model}

% 窗口主题模型介绍
The window topic model projects documents into a topic space using the NMF method. The loss function for the document collection $Z^{t}$ in time period $t$ within the window topic model is defined as follows:

% 第一层窗口主题模型公式：
\begin{equation}
\begin{array}{ll}
\min & L_1^t\left(X^{t}, Y^{t}\right)=\left\|Z^{t}-X^{t} Y^{t}\right\|_F^2, \\
& t=1, \ldots, T, \\
\mathrm{s.t.} & X^{t} \geq 0, \quad Y^{t} \geq 0 .
\end{array}
\end{equation}

% 第一层窗口主题模型解释：
\noindent where the text matrix $Z^{t}$ represents the document collection for time period $t$ ($t \in [1, \ldots, T]$). When the document corpus is divided into $T$ time periods, $T$ window topic models must be trained. Each row of $Z^{t}$ represents a document vector, and each column corresponds to a word vector. The matrix $X^{t}$ serves as the window topic matrix, capturing the weights associated with different topics present in the text data for each time period. Meanwhile, $Y^{t}$ is the window feature matrix, which stores feature weight vectors for each topic. Consequently, the window topic model for the document collection $Z^{t}$ in the DNMF method is represented by the window topic matrix $X^{t}$ and the window feature matrix $Y^{t}$.

\subsubsection{Dynamic Topic Model}

% 动态主题模型介绍
The window topic model generates $T$ topic models, each with a distinct window feature matrix, as their topic feature spaces may vary. Therefore, to align feature spaces, the window topic term matrices are integrated into a reconstructed term feature space. By aggregating the $T$ window feature matrices $Y^{t}$, a reconstructed window feature matrix $V=\left[Y^{1}, \ldots, Y^{T}\right]$ is obtained, where $V^{t}$ represents the reconstructed window feature matrix for document collection $Z^{t}$ during time period $t$.

In the second layer of the DNMF model, window topic features across time periods are integrated using a moving-window approach, enabling the extraction of topic evolution through the NMF model. The loss function for the dynamic topic model is defined as:

% 第二层动态主题模型公式：
\begin{equation}
\begin{array}{ll}
\min & L_2(W,H)=\left\|V-WH\right\|_F^2, \\
\mathrm{ s.t. } & W \geq 0, \quad H \geq 0 .
\end{array}
\end{equation}

% 第二层动态主题模型解释：
\noindent where the matrix $H$ represents the dynamic feature matrix, playing a dual role in the second layer. First, it acts as the topic feature matrix for the reconstructed window feature matrix $V$. Second, it serves as the dynamic feature matrix for each document collection $Z^{t}$. Each row vector in $H$ represents a dynamic topic, with the top-ranked feature words serving as the topic's keywords. The matrix $W$ is referred to as the reconstructed topic matrix, representing the topic matrix for the reconstructed window feature matrix $V$.

The modeling process in the second layer is different from that of the first layer as it reconstructs the window feature matrices $H^{t}$ for documents across different time periods. Consequently, the dynamic topic matrix $F^{t}$ for each corpus $H^{t}$ is computed. The dynamic topic matrix for document collection $Z^{t}$ is expressed as:
% 动态主题权重矩阵表达式：
\begin{equation} \label{eq:dywv}
\begin{array}{ll}
Z^{t}& \approx X^{t}Y^{t}  \\
&=X^{t}V^t \\
&\approx X^{t}W^{t}H \\
&=F^{t}H
\end{array}
\end{equation}

\noindent where $F^{t}=X^{t}W^{t}$ represents the dynamic topic matrix of the document collection $Z^{t}$. The weights in the row vectors of the reconstructed topic matrix $W^{t}$ indicate how relevant each window topic is to the final dynamic topic. The equality $V^t=Y^{t}$ holds due to the reconstructed window feature matrix $V$ integrating the window feature matrix $Y^{t}$ accordingly.

Eq. \ref{eq:dywv} reveals that the window topic and the reconstruction topic matrix influence the optimization of the dynamic topic matrix. For sparse text data, it is essential to incorporate the problem's requirements and text sparsity to optimally reconstruct the topic matrix.

\subsection{Stage 2: Dynamic Topic Refinement}

% \begin{figure}[!t]
% \centering
% \includegraphics[width=\linewidth]{fig/fig5.png}
% \caption{Dynamic convex NMF}
% \label{fig:dcnmf}
% \end{figure}

% \documentclass{article}
% \usepackage{graphicx}

% \begin{figure*}[!t]
%     \centering
%     \resizebox{\textwidth}{\textheight}{\includegraphics{fig/fig5.png}} % 或使用 \textheight - \baselineskip 来稍微减小一些高度
%     \caption{Dynamic convex NMF}
% \end{figure*}
\begin{figure}[!t]  % 't' 选项尝试将图形置于页面的顶部
    \centering
    \includegraphics[width=7.6\textwidth,height=22.8cm,keepaspectratio]{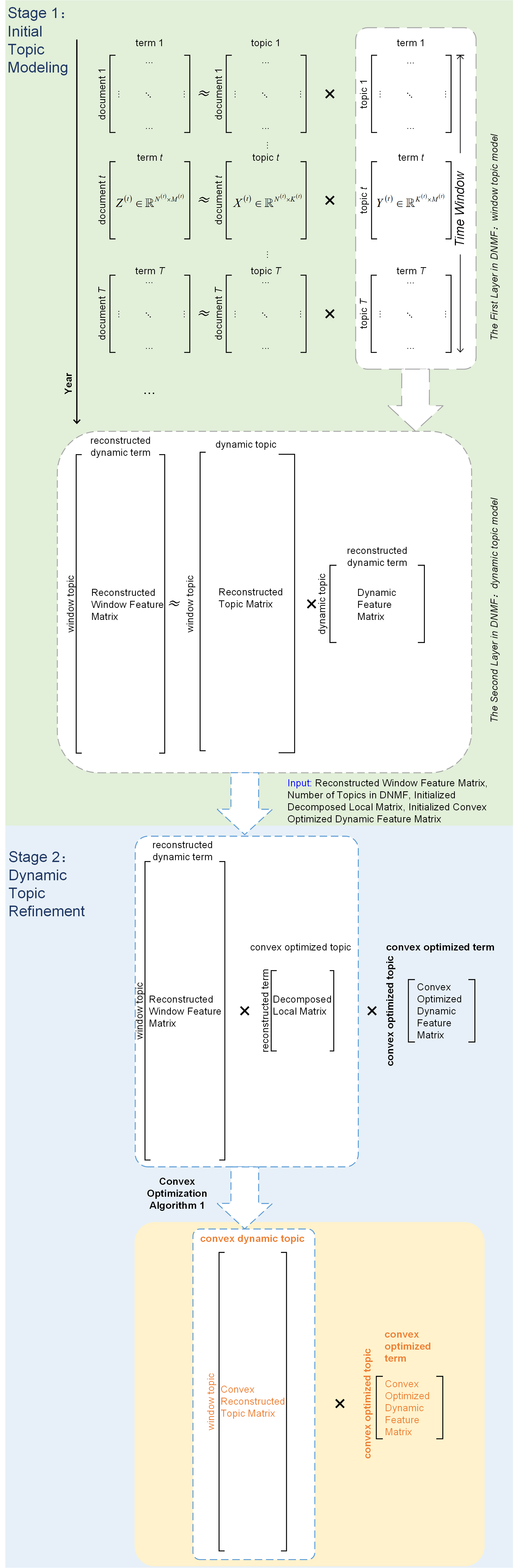}
    \caption{Dynamic convex NMF}
    \label{fig:dcnmf}
\end{figure}

% \begin{figure}[!t]
%     \centering   \includegraphics[width=\textwidth,height=\textheight,keepaspectratio]{fig/fig5.png}
%     \caption{Dynamic convex NMF}
% \end{figure}

The flowchart of the proposed two-stage dynamic topic modeling approach is shown in Fig. \ref{fig:dcnmf}. Compared with the two-layer NMF-NMF model, this section enhances the dynamic topic model using convex non-negative matrix factorization.

The NMF model combines the strengths of clustering algorithms and mathematical programming\cite{recht2012NMFlp}. This integration allows for further optimization of its results in topic analysis tasks. The second layer of DNMF models exhibits both clustering capabilities and mathematical modeling characteristics. To improve the effectiveness, sparsity, and stability of online text topic mining, we incorporate a convex optimization technique into the DNMF model. This technique further optimizes the reconstructed topic matrix $W$, dynamic topic matrix $F$, and dynamic feature matrix $H$. Additionally, the topic number $K$ and dynamic feature matrix $H$ from the DNMF model serve as initial values for hyperparameter fine-tuning.

The convex NMF model extends traditional non-negative matrix factorization, providing robust solutions for handling noise and outliers, thereby enhancing the accuracy and robustness of text topic analysis. The convex NMF model, as a text mining technique, automatically extracts latent topics from large-scale text data, representing them as the product of a topic distribution matrix and a word distribution matrix. This model facilitates researchers in comprehending the meaning and structure of text data, uncovering hidden topic structures within text data, and providing support for applications such as text classification, information retrieval, and recommendation systems.

% cNMF模型中W*,H*的初始化
The convex NMF model is used for the refinement of the dynamic topic model. We approximate the reconstructed topic matrix $W$ of the DNMF model and the matrices $G$ as a convex linear combination of matrix $V$, as follows.

\begin{equation}
V \cdot G_{k}^{*} = W_{k}, k=1,...,K
\end{equation}

\noindent where $G=\left(G_{1},\ldots G_{K}\right)$ is called the decomposed local matrix. $W$ is the reconstructed topic matrix from the DNMF model in stage 1. 

The convex NMF optimization algorithm typically uses clustering algorithms, such as k-means, to initialize matrices $G$ and $\tilde{H}$, followed by updating matrices $\tilde{W}$ and $\tilde{H}$ using the multiplicative update rule. Here, the dynamic feature matrix $H$ of the DNMF model is initialized as the dynamic feature matrix $\tilde{H}$ in the convex NMF.

Based on the topic clustering results of the DNMF model, the reconstructed topic matrix $W=\left(W_{1},...W_{K}\right)$ is divided into $K$ classes. Therefore, the initialization of the matrix $G$ is developed concerning a linear programming model as follows.

\begin{equation}
\label{eq:decomposed local matrix}
\begin{array}{ll}
\min&\sum_{k=1}^{K}\|W_{k}-V \cdot {G_{k}}^{*}\|_{F}^{2},\\
\mathrm{s.t.} & G\geq0.
\end{array}
\end{equation}

Based on the convex non-negative matrix factorization model \cite{ding2008convex}, a topic refinement model is presented for dynamic topic results as follows.

% cNMF模型
\begin{equation}
\begin{array}{ll}
\min & L_3(G,\tilde{H})=\frac{1}{2}\|V-VG\tilde{H}\|_F^2\\
\mathrm{s.t.} & G \geq 0, \quad \tilde{H} \geq 0.
\end{array}
\end{equation}

\noindent where $\tilde{H}$ represents the convex optimized dynamic feature matrix $H$, $G$ is the decomposed local matrix considering the Eq. \ref{eq:decomposed local matrix}, and then the convex optimized reconstructed topic matrix $\tilde{W}=VG$.

The pseudocode for the optimization algorithm proposed for the convex NMF problem in Stage 2 is as follows:

% \begin{algorithm}[H]
\begin{algorithm}
\caption{Pseudocode for Convex NMF in Dynamic Topic Modeling}
\label{alg:pseudocode}
\DontPrintSemicolon % Some LaTeX compilers require you to use\dontprintsemicolon instead
\KwIn{Data matrix $V \in \mathbb{R}^{n*m}$, number of topics $K$ (number of basis vector),  initial matrix $G_{init}$,  initial matrix $H_{init}$,iteration count $I$ and threshold $\delta$}
\KwOut{optimized matrices: $\tilde{H}$, $G$ and $\tilde{W}$}

Initialize $G$ and $\tilde{H}$:  $G=G_{init}$ and $\tilde{H}=H_{init}$ 

Calculate the initial matrix $\tilde{W}$: $\tilde{W}=VG$

Update the matrices $G$, $\tilde{W}$ and $\tilde{H}$:\\
\For{$i=1,...,I$}
{a. Update the decomposed local matrix $G$ and the reconstructed topic matrix $\tilde{W}$:\\
$G\leftarrow G\sqrt{\frac{\left[\left(V^T\cdot V\right)^+\cdot H^{t}\right]^++\left(V^T\cdot V\right)^-\cdot G\cdot H\cdot H^{t}}{\left[\left(V^T\cdot V\right)^-\cdot H^{t}\right]^-+\left(V^T\cdot V\right)^+\cdot G\cdot H\cdot H^{t}}}$

$\;\tilde{W}\leftarrow \tilde{W}\cdot G$

b. Update the dynamic feature matrix $\tilde{H}$:\\
$\tilde{H}=\tilde{H}\frac{\sqrt{\left(G\right)^T \cdot \left(\left(V^T \cdot V\right)^+ +\left(V^T \cdot V\right)^- \cdot G \cdot \tilde{H}\right)}}{\tilde{H}\left(\left(V^T \cdot V\right)^- +\left(V^T \cdot V\right)^+\cdot G \cdot \tilde{H}\right)}$

c. Calculate the loss function value for the current iteration:\\
$L=\|V-\tilde{W}\tilde{H}\|_F^2$

d. If the change in the loss function is less than the threshold, stop the iteration:\\
  \If{$L<\delta$}
  {
    \textbf{break};
  }
}
\Return{$G$, $\tilde{W}$ and $\tilde{H}$}
\end{algorithm}

The positive and negative parts of a matrix $A$ are defined as:

\begin{equation}
(A^{T} \cdot A)^{+} = \frac{1}{2}(|A^{T} \cdot A|+A^{T} \cdot A)
\end{equation}
\begin{equation}
(A^{T} \cdot A)^{-} = \frac{1}{2}(|A^{T} \cdot A|-A^{T} \cdot A)
\end{equation}

The convex non-negative matrix factorization model can be further extended to include a regularization term constraint, as studied by Kim and Park\cite{kim2008nonnegative}, and a sparsity constraint, as studied by Hoyer \cite{hoyer2004non}. This extension enables the NMF-based topic mining and analysis method to be applied to various scenarios. These extended convex non-negative matrix factorization models can be solved using convex optimization techniques, further improving the accuracy and adaptability of the model.

Moreover, the two-stage approach supports both offline and online analysis paradigms: Stage 1 (Offline Topic Modeling) focuses on offline modeling of large-scale data, while Stage 2 (Online Dynamic Topic Refinement) enables real-time dynamic topic refinement. Together, these stages provide a cohesive and powerful methodology for comprehensive topic modeling. In practice, if the first stage is designed for offline training, the initialization of the topic matrix and feature matrix of the convex NMF model can also be implemented in the first stage.

\section{Experimental Evaluation}
\subsection{Dataset Description}

According to the official introduction, the IEEE T-ASE journal covers the following research areas:
\begin{itemize}
    \item Basic research on automation technology.
    \item Applications of automation technology in the fields of power, energy, and industrial applications.
    \item How automation technology can improve efficiency, quality, productivity, and reliability in industrial applications.
\end{itemize}

\begin{table}[!htbp]
\caption{Research topics of T-ASE journal\label{tab:research-topics-TASE}}
\centering
\begin{tabular}{lcl}
\hline
\multicolumn{3}{l}{T-ASE journal}                                                                                                                                                                                                                     \\ \hline
\multirow{3}{*}{Discipline} & 1 & Components, Circuits, Devices and Systems                                                                                                                                                                           \\
                            & 2 & Power, Energy and Industry Applications                                                                                                                                                                             \\
                            & 3 & Robotics and Control Systems                                                                                                                                                                                        \\ \hline
\multirow{6}{*}{Scope}      & 1 & \begin{tabular}[c]{@{}l@{}}computer science, control systems, electrical\\ engineering, mathematics, mechanical engineering,\\ operations research, and other fields\end{tabular}                                   \\
                            & 2 & \begin{tabular}[c]{@{}l@{}}agriculture, biotechnology, healthcare, home \\ automation, maintenance, manufacturing, \\ pharmaceuticals, retail, security, service, \\ supply chains, and transportation\end{tabular} \\
                            & 3 & efficiency, quality, productivity, and reliability                                                                                                                                                                  \\ \hline
\end{tabular}
\end{table}

The dataset for this study consists of abstract data from articles published in the IEEE T-ASE journal, including content from the "Note to Practitioners" section. Spanning from the inaugural issue in January 2004 to the second issue of 2022, the dataset encompasses 19 volumes in total. A lifetime topic analysis was conducted on the abstracts of all articles published during this period, with a dynamic topic analysis focusing on abstracts from 2020 to 2022. To strike a balance between topic diversity and model interpretability, the optimized range for the number of topics was set at 5 to 20.

\subsection{Experiment 1: Selection of Topic Models}
This study conducted topic analysis on the full-cycle abstract dataset of the T-ASE journal from 2004 to 2022, providing both the window topic mining results for each yearly abstract dataset and the dynamic topic mining results for the full-cycle abstract dataset.

NMF and LDA stand out as two of the most representative unsupervised topic models. However, these models differ in terms of their optimization techniques. The NMF model emphasizes matrix constraints and sparsity, while the LDA model focuses on probabilistic modeling of topics and vocabulary. Considering that journal abstracts are concise and brief text data, in Experiment 1, we utilized the NMF model and the LDA model to conduct topic mining on the text datasets of the full-cycle abstracts. This allowed us to compare and analyze the influence of the abstract dataset on the adaptability of the models.

Table 5.2 presents the results of topic analysis on the full-cycle abstracts using the NMF and LDA models with different numbers of topics (i.e., 6, 12, 18, and 24). To further compare the performance of topic models on the yearly abstract data, the coherence ($C_V$)\cite{roder2015exploring} and uniformity ($C_Umass$)\cite{newman2010automatic} metrics were evaluated for both NMF and LDA models. The results, with the number of topics set to 6 and 18, are presented in Table \ref{tab:cv-window-topic-models} and Table \ref{tab:umass-window-topic-models}, respectively.

\begin{table}[!htbp]
\caption{Lifetime Topic Modeling Evaluation\label{tab:nmf&lda}}
\centering
\begin{tabular}{cc|cccc}
\hline
\multicolumn{2}{c|}{Topic Number} & \begin{tabular}[c]{@{}c@{}}Training\\ Time (s)\end{tabular} & TC-W2V & CV    & CUMass \\ \hline
\multirow{2}{*}{K=6}     & NMF   & 0.161                                                       & 0.226  & 0.260 & -0.635 \\
                         & LDA   & 4.411                                                       & 0.163  & 0.250 & -0.550 \\ \cline{1-2}
\multirow{2}{*}{K=12}    & NMF   & 0.699                                                       & 0.171  & 0.260 & -1.166 \\
                         & LDA   & 4.766                                                       & 0.102  & 0.255 & -0.659 \\ \cline{1-2}
\multirow{2}{*}{K=18}    & NMF   & 0.734                                                       & 0.117  & 0.266 & -1.412 \\
                         & LDA   & 4.566                                                       & 0.092  & 0.256 & -0.630 \\ \cline{1-2}
\multirow{2}{*}{K=24}    & NMF   & 0.692                                                       & 0.093  & 0.263 & -2.275 \\
                         & LDA   & 5.024                                                       & 0.070  & 0.256 & -0.697 \\ \hline
\end{tabular}
\end{table}

From the results in Table \ref{tab:nmf&lda}, it can be observed that the NMF model outperforms the LDA model in terms of topic clustering for the full-cycle abstract text dataset. By comparing training time, reconstruction error, and topic consistency, the NMF model proves to be more suitable than LDA as the offline training method for Stage 1 in the proposed two-stage dynamic topic modeling framework. Particularly, the training time and reconstruction error show significant improvements compared to the LDA model, enabling efficient distributed feature extraction of the abstract dataset across multiple periods.

\begin{table}[!htbp]
\caption{CV of Window Topic Models\label{tab:cv-window-topic-models}}
\centering
\begin{tabular}{ccccc}
\hline
Cv   & NMF            & LDA            & NMF             & LDA    \\
Year & \multicolumn{2}{c}{K=6}         & \multicolumn{2}{c}{K=18} \\ \hline
2004 & 0.316          & 0.270           & 0.448  & 0.32   \\
2005 & 0.278          & 0.295 & 0.323  & 0.312  \\
2006 & 0.284          & 0.293 & 0.328  & 0.283  \\
2007 & 0.300          & 0.256          & 0.452  & 0.312  \\
2008 & 0.288          & 0.288          & 0.388  & 0.307  \\
2009 & 0.319          & 0.294          & 0.342  & 0.315  \\
2010 & 0.287          & 0.303 & 0.333  & 0.291  \\
2011 & 0.285          & 0.259          & 0.379  & 0.305  \\
2012 & 0.282          & 0.27           & 0.367  & 0.278  \\
2013 & 0.299          & 0.283          & 0.386  & 0.302  \\
2014 & 0.292          & 0.308 & 0.307  & 0.3    \\
2015 & 0.320          & 0.280           & 0.345  & 0.284  \\
2016 & 0.257          & 0.290  & 0.320   & 0.295  \\
2017 & 0.289          & 0.291 & 0.310   & 0.289  \\
2018 & 0.281          & 0.285 & 0.305  & 0.281  \\
2019 & 0.269          & 0.283 & 0.375  & 0.281  \\
2020 & 0.274          & 0.289 & 0.301  & 0.283  \\
2021 & 0.377          & 0.254          & 0.340   & 0.287  \\
2022 & 0.356          & 0.268          & 0.363  & 0.273  \\ \hline
\end{tabular}
\end{table}

\begin{table}[!htbp]
\caption{CV of Window Topic Models\label{tab:umass-window-topic-models}}
\centering
\begin{tabular}{ccccc}
\hline
Umass & NMF             & LDA    & NMF             & LDA    \\
Year  & \multicolumn{2}{c}{K=6}  & \multicolumn{2}{c}{K=18} \\ \hline
2004  & -2.565 & -1.334 & -5.216 & -1.300   \\
2005  & -1.869 & -0.609 & -3.928 & -1.092 \\
2006  & -2.042 & -0.626 & -3.255 & -1.043 \\
2007  & -1.762 & -0.744 & -3.224 & -1.23  \\
2008  & -1.688 & -0.572 & -4.271 & -0.606 \\
2009  & -1.299 & -0.449 & -2.761 & -0.833 \\
2010  & -0.815 & -0.438 & -3.788 & -0.793 \\
2011  & -3.008 & -0.572 & -3.540 & -0.719 \\
2012  & -1.413 & -0.579 & -2.465 & -1.073 \\
2013  & -1.297 & -0.514 & -2.196 & -0.621 \\
2014  & -1.589 & -0.465 & -3.195 & -0.873 \\
2015  & -0.950 & -0.376 & -4.786 & -0.600   \\
2016  & -1.010 & -0.432 & -2.859 & -0.714 \\
2017  & -1.063 & -0.372 & -2.560 & -0.669 \\
2018  & -0.764 & -0.344 & -2.826 & -0.634 \\
2019  & -1.242 & -0.322 & -5.582 & -0.536 \\
2020  & -1.301 & -0.269 & -2.172 & -0.626 \\
2021  & -0.788 & -0.288 & -2.088 & -0.832 \\
2022  & -0.588 & -0.310 & -1.790 & -0.392 \\ \hline
\end{tabular}
\end{table}

In terms of the analysis of topic coherence for the yearly abstract dataset, the NMF model significantly outperforms the LDA model in terms of the $C_Umass$ topic coherence metric across different periods and different numbers of topics. As for the $C_V$ topic coherence metric, when the number of topics $K$ is set to 18, the topic coherence of the NMF model's clustering results is consistently better than that of the LDA model. However, when the number of topics is relatively small, the topic coherence of the NMF model's clustering results is slightly weaker than that of the LDA model in a few specific years.

For datasets such as journal abstracts, which are characterized by short texts, small sample sizes, and a relatively large number of topics, the NMF model performs better than the LDA model in terms of topic clustering. Additionally, when considering dynamic topic clustering, the NMF model is more suitable than the LDA model as a window topic model.

\begin{figure}[!t]
\centering
\includegraphics[width=\linewidth]{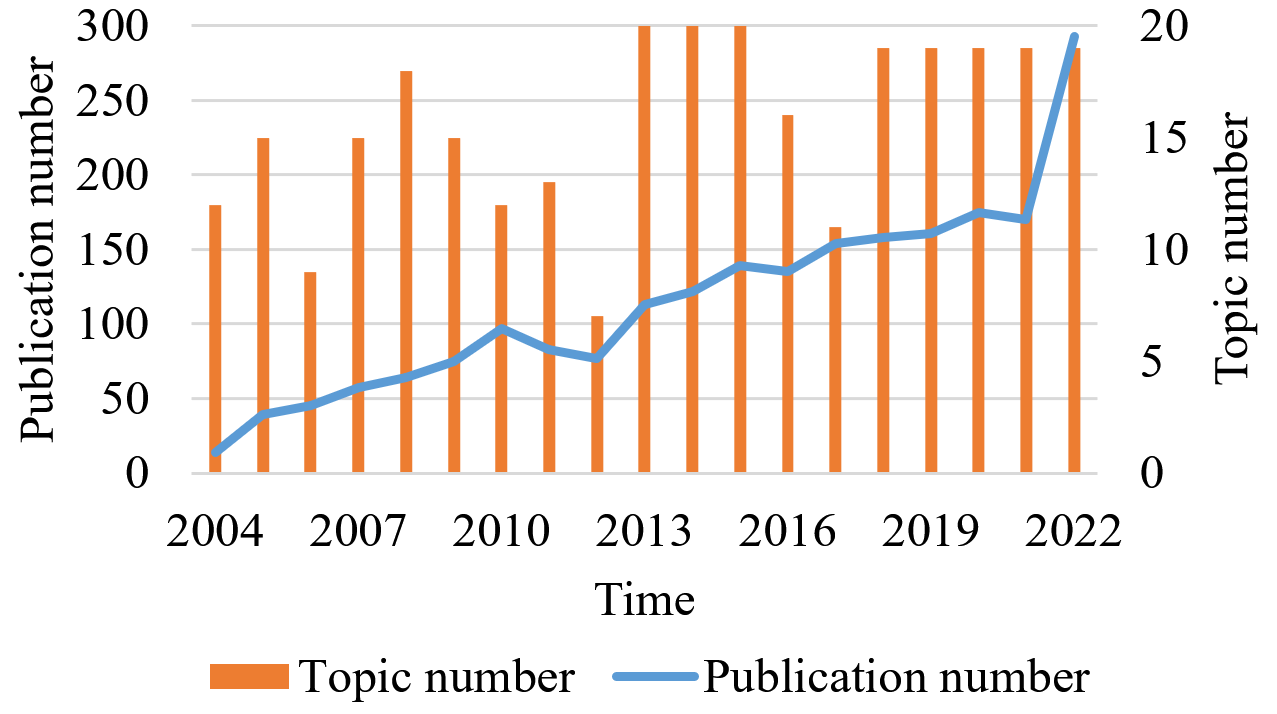}
\caption{Topic number selection of dynamic topic model based on TC-W2V score}
\label{fig:topic&publication}
\end{figure}

From the topic clustering results of the NMF-NMF model in Fig. \ref{fig:topic&publication}, it can be observed that although the number of published journal articles shows an increasing trend over the years, there is significant fluctuation in the number of topics across different years. However, when looking at the timeline, the variation in the number of topics appears to exhibit a potential periodicity. The years 2008, 2013, and 2019 are identified as peak years. From 2018 to 2022, the optimal number of topics remains consistently high. This can be attributed to the increase in the number and content of articles, but it also indicates the problem of overfitting when the number of articles far exceeds the number of topics.

Next, the full-cycle dynamic topic analysis of the T-ASE journal is presented. Based on the TC-W2V topic coherence measure, the optimal number of topics obtained from the dynamic topic model is 18. The following figure shows the TC-W2V change curve for the DNMF model under different numbers of topics.

Lastly, to validate the performance of the convex non-negative matrix factorization (cNMF) model in dynamic topic mining, this study conducted NMF-NMF and NMF-cNMF models based on the recent abstract data from 2020 to 2022. Fig. \ref{fig:topic&publication} shows that for the TC-W2V optimized NMF window topic model, the optimal number of topics for 2020, 2021, and 2022 is 19. During these three years, to ensure topic diversity and representativeness, four topics were selected for discussion: robotics, path planning (autonomous driving), COVID-19, and digital twin. According to the author's knowledge, these four topics have been popular research areas since 2020. Based on the analysis of the overall dynamic and window topic models, robotics and path planning are considered classic research topics in the T-ASE journal, while COVID-19 and digital twin represent emerging research topics after 2020.

\begin{figure}[!t]
\centering
\includegraphics[width=\linewidth]{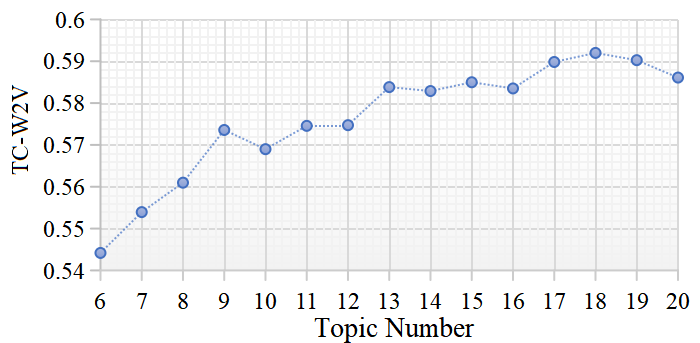}
\caption{Topic number selection of dynamic topic model based on TC-W2V score}
\label{fig:dnmf_coherence}
\end{figure}

\subsection{Experiment 2: Lifetime Topic Analysis}
\subsubsection{Results analysis of the window topic model}
The results of the DNMF topic model reveal significant fluctuations in the number of topics across different years, with notable peaks observed in 2008, 2013, and 2019. From 2018 to 2022, the optimal number of topics remains consistently high. This trend can be attributed not only to the growing volume and diversity of articles but also to potential overfitting issues arising from the disproportionate relationship between the number of articles and topics.

Analyzing and predicting yearly topic trends presents a considerable challenge for human scholars and editors. Furthermore, the rapid emergence of numerous new topics creates obstacles for leveraging academic journals effectively in scientific research and industrial development.

\subsubsection{Results analysis of the dynamic topic model}
The journal topic clustering or list includes the number of topics, topic categories, importance ranking, and specific features under each topic. For journals such as IEEE T-ASE, which involve interdisciplinary research in industrial fields, providing a satisfactory journal topic list is quite challenging. Therefore, this paper first uses the DNMF method to provide a baseline result for topic clustering analysis and manual double-checking of the T-ASE journal. Thus, based on the baseline result, a subjectively feasible number of topics and approximate topic clustering and keywords are defined.
Next, a full-cycle topic analysis of the T-ASE journal is presented. Based on TC-W2V topic consistency, the optimal number of topics obtained by the dynamic topic model is 18. The popularity ranking of topics and high proportion keywords in each topic are shown in Table \ref{tab:2020-2022-Topics-NMF-cNMF}. Fig. \ref{fig:dnmf_coherence} shows the TC-W2V change curve corresponding to the DNMF model under different topic settings.

\begin{table*}[!htbp]
\caption{Topics of NMF-NMF model and NMF-cNMF model during 2020-2022 (K=19)\label{tab:2020-2022-Topics-NMF-cNMF}}
\centering
\begin{tabular}{llll}
\hline
Topic                                                                    & Model    & Rank & Top 20 words                                                                                                                                                                                                                                                                                                                                                           \\ \hline
\multirow{2}{*}{Robot}                                                   & NMF-cNMF & 1    & \begin{tabular}[c]{@{}l@{}}robot, human, task, human-robot, interaction, skill, mobile, mobile robot, motion, robot human, environment,\\ physical, learn, enable, robotic, end, localization, manipulator, intelligent, formation\end{tabular}                                                                                                                        \\
                                                                         & NMF-NMF  & 9    & \begin{tabular}[c]{@{}l@{}}robot, human, task, human-robot, interaction, skill, motion, robot human, learn, robotic, physical, \\ environment, assembly, enable, end, hand, intelligent, gait, intention, formation\end{tabular}                                                                                                                                       \\ \hline
\multirow{2}{*}{\begin{tabular}[c]{@{}l@{}}Path\\ planning\end{tabular}} & NMF-cNMF & 3    & \begin{tabular}[c]{@{}l@{}}path, planning, trajectory, path planning, motion, vehicle, length, plan, coverage, sensor, trajectory\\ planning, obstacle, autonomous, UGV (Unmanned Ground Vehicle), simulation, track, generate, follow, area,\\ complex\end{tabular}                                                                                                    \\
                                                                         & NMF-NMF  & 8    & \begin{tabular}[c]{@{}l@{}}path, planning, path planning, trajectory, motion, vehicle, plan, length, coverage, sensor, autonomous, \\ trajectory planning, generate, UGV, simulation, obstacle, robot, complex, area, follow\end{tabular}                                                                                                                              \\ \hline
\multirow{2}{*}{Covid}                                                   & NMF-cNMF & 18   & \begin{tabular}[c]{@{}l@{}}covid, cosmos, cc19lp, pandemic, illness, fatality, county, risk, high risk, cosmos covid, community,\\ population, former, intervention, face, latter, group, covid pandemic, individual, minimize\end{tabular}                                                                                                                            \\
                                                                         & NMF-NMF  & 15   & \begin{tabular}[c]{@{}l@{}}covid, cosmos, cc19lp, pandemic, county, fatality, illness, DHHFSP- NTOU, DHHFSP, NTOU, individual, aco\_moea,\\ risk, electricity, cosmos covid, high risk, population, community, former, intervention\end{tabular}                                                                                                                        \\ \hline
\multirow{2}{*}{\begin{tabular}[c]{@{}l@{}}Digital\\ twin\end{tabular}}  & NMF-cNMF & 19   & \begin{tabular}[c]{@{}l@{}}digital twin, DTIM(digital twin intelligence manufacturing), manufacturing, industry, technology, intelligent,\\ conveyor, manufacturing industry, twin, intelligence digital twin, product digital twin, physical,\\ intelligent manufacturing, product, intelligence, twin digital twin, virtual, service, integration, core\end{tabular} \\
                                                                         & NMF-NMF  & 17   & \begin{tabular}[c]{@{}l@{}}digital twin, DTIM, manufacturing, industry, intelligent, technology, manufacturing industry, conveyor, twin,\\ product digital twin, intelligence digital twin, physical, intelligent manufacturing, product, intelligence,\\ service, twin digital twin, achieve, virtual, core\end{tabular}                                              \\ \hline
\end{tabular}
\end{table*}

\subsection{Experiment 3: Dynamic Topic Analysis}
To verify the performance of the convex non-negative matrix factorization model in dynamic topic mining, this paper performs NMF-NMF and NMF-cNMF models based on the latest abstract data from 2020 to 2022. To ensure the diversity and representativeness of topic mining for the three years from 2020 to 2022, we selected four topics, namely robots, path planning (autonomous driving), COVID-19, and digital twins, as discussion objects. According to the author's knowledge, these four topics became popular after 2020. From the results analysis of the full-cycle and window topic models, robots and path planning belong to classic research topics in the T-ASE journal, while COVID-19 and digital twins belong to emerging research topics after 2020.

Analysis of Window Topic Models for 2020, 2021, and 2022 reveals that the optimal number of topics ($K$) for the NMF model is 19. In the topic clustering results of the window topic model for 2020, the topic most related to robots is the human-machine interaction topic, ranking second, while the path planning topic ranks eighth, and the digital twins and COVID-19 topics have no ranking. In the topic clustering results for 2021, the robot topic ranks first, but the top 20 words in this topic do not include "motion" or words related to robot motion control. The path planning topic ranks 13th, while the digital twins and COVID-19 topics have no ranking. In the topic clustering results for 2022, the robot topic ranks first, the path planning topic ranks second, the digital twins topic ranks 15th, and the COVID-19 topic ranks 16th.

Dynamic topic mining was performed on the window topic models for 2020-2022, and the experimental results are shown in Table \ref{tab:2020-2022-Topics-NMF-cNMF}. It shows the four topic clustering results of robots, path planning (autonomous driving), COVID-19, and digital twins, and analyzes the top 20 keyword rankings for each topic.

For emerging popular topics, the dynamic topic mining effect of the NMF-cNMF model is compared with that of the NMF-NMF model. The proposed NMF-cNMF model successfully integrates the two new emerging topics, "digital twins" and "COVID-19," which only appeared in the window topic clustering results for 2022, and their keywords into the topic model.

In the experimental results of the NMF-NMF model, the robot topic ranks low because the same topic clustering method as the window NMF model is used, and robots are grouped under the human-machine interaction category. The NMF-cNMF algorithm integrates human-machine interaction and robot motion into a new topic category and ranks robots first.

For classic popular topics, the dynamic topic mining effect of the NMF-cNMF model is compared with that of the NMF-NMF model. On the one hand, regarding the robot topic, in the experimental results of the NMF-NMF model, the robot topic ranks low because the same topic model as the window NMF model is used, and the topic clustering result also places robots as keywords under the human-machine interaction category, rather than integrating human-machine interaction and robot motion and control into the robot topic. The proposed NMF-cNMF model integrates human-machine interaction and robot motion and control into a new topic category (robots) and ranks it first. On the other hand, for the path planning (autonomous driving) topic, the order and content of the topic keywords in the NMF-NMF model and the NMF-cNMF model are similar. However, in the results of the NMF-cNMF model, the path planning (autonomous driving) topic's ranking has risen from 8th to 3rd. Considering sparsity, optimality, and topic model evolution, that is, the ranking of the path planning topic in the window model has evolved from 8th in 2020 to 13th in 2021 and 2nd in 2022. The NMF-cNMF model shows better performance in tracking topic evolution trends compared to the NMF-NMF model.

Furthermore, by considering the sparsity among feature words, the experimental results reveal the extent of ranking changes between the two models for emerging and traditional topics. These changes reflect the degree of association between emerging and traditional topics, as well as among traditional topics. A greater magnitude of change indicates a stronger degree of association.

\subsection{Experiment 4: Model Stability}
To verify the stability of the proposed convex non-negative matrix factorization method, the topic clustering results of the dynamic topic model are comparatively analyzed by adding different tiny sparse term penalties to the window topic model. The expression of the sparse NMF(sNMF) model is as follows:

\begin{equation}
\begin{array}{ll}
L_{\mathrm{l}}\left(W,\mathrm{H}\right)& ={\frac{1}{2}}\|V-W\mathrm{H}\|_{F}^{2}  \\
&+\alpha\cdot l\cdot n_{1}\left\|W\right\|_{1}+\beta\cdot l\cdot n_{2}\left\|H\right\|_{1} \\
&+\frac{1}{2}\alpha\cdot\left(1-l\right)\cdot n_{1}\left\|W\right\|_{F}^{2} \\
&+\frac{1}{2}\beta\cdot\left(1-l\right)\cdot n_{2}\left\|H\right\|_{F}^{2}
\end{array}
\end{equation}

\noindent where $n_{1}$ and $n_{2}$ represent the number of text features and samples, respectively. The parameters $\alpha$, $\beta$ and $l$ are sparse regularization terms for the NMF model. 

Topic consistency based on the Word2Vec model (TC-W2V) is used to assess the topic mining and parsing effectiveness of the dynamic topic clustering models sNMF-NMF and sNMF-cNMF under the effect of different sparsity constraints of the window model. In addition, the sum of the average weighted absolute values of topic ranking changes compared to NMF-NMF and NMF-cNMF, respectively, was used to assess the stability of the two dynamic topic models. The degree of topic ranking change is defined as follows:

\begin{equation}
\frac{1}{K}\sum_{k=1}^{K}\big|pop_{ra(DT_{k})}{}\times ra(DT_{k})-pop_{k}{} \times k\big|
\end{equation}

\noindent where $DT_k$ represents the $kth$ dynamic topic without sparsity constraints and $ra(DT_{k})$ represents the new ranking of topic $DT_k$ with sparsity constraints. $pop_k$ is the popularity share of the $kth$ dynamic topic. It can be seen that the degree of change of topic ranking is positive, when it is smaller, it means that the change of the product of topic ranking and popularity ratio is smaller, and the result of topic mining is relatively stable.

let $\alpha=0.00001$, $\beta=0$, $K=18$. A tiny sparse perturbation is introduced to the first layer of the NMF-based window topic model by adjusting the regular term $l$, which is used to analyze the stability of the topic ranking of the second layer of the dynamic topic model as follows.

\begin{table}[!htbp]
\caption{Model Stability (K=18)\label{tab:model-stability}}
\centering
\begin{tabular}{ccccc}
\hline
\multirow{2}{*}{$l$} & \multicolumn{2}{c}{TC-W2V} & \multicolumn{2}{c}{Change in rankings of topics} \\
                   & sNMF-NMF    & sNMF-cNMF    & sNMF-NMF               & sNMF-cNMF               \\ \hline
0                  & 0.5585      & 0.5622       & 0                      & 0                       \\
0.1                & 0.5559      & 0.5535       & 0.0912                 & 0.0938                  \\
0.4                & 0.559       & 0.552        & 0.0926                 & 0.0699                  \\
0.6                & 0.5522      & 0.5634       & 0.1484                 & 0.0644                  \\
0.9                & 0.5548      & 0.5419       & 0.1614                 & 0.1018                  \\ \hline
\end{tabular}
\end{table}

The TC-W2V values of the NMF-NMF and NMF-cNMF models without sparsity constraints are demonstrated in Table 5.7 as 0.5585 and 0.5622, respectively, when the number of topics is set to 18. At this point, the cNMF model slightly improves the topic consistency by about 0.66\%. When the sparsity of the first layer sNMF model increases from 0.1 to 0.9, the TC-W2V values and the degree of change of topic ranking of the second layer NMF and cNMF models. It can be seen that when the sparsity of the first layer sNMF model is gradually increased, there is no significant change in the TC-W2V value of cNMF compared to NMF. Moreover, when the sparsity of the first layer sNMF model is small (sparseness=0.1), the subject ranking variation of cNMF compared to NMF is similar. When the sparsity of the first layer of the sNMF model gradually increases, the degree of topic ranking variation of cNMF compared to NMF decreases significantly, indicating that cNMF is less affected by the sparsification of the feature word space from the first layer of the window topic model. Specifically, at sparsity levels of 0.4, 0.6, and 0.9, the relative reduction in topic ranking variability is 24.51\%, 56.60\%, and 36.93\%, resulting in a more stable topic ranking.

From the topic clustering results of the benchmark experiment, the NMF-cNMF structure is more stable compared to the NMF-NMF structure. Whether it is full-cycle topic analysis or recent topic analysis, the NMF-cNMF model's topic ordering and keywords and the relationship between them present more regular topics and features, and the validity of the topic clustering results is stronger.

\subsection{Discussion}
The NMF used for dynamic topic analysis is a word-frequency-based method. Therefore, the topic ranking in dynamic topic analysis tasks is easily disturbed by word frequency. If there are many low-frequency words or phrases in the text dataset, or if the dataset is unbalanced, then the model fitting optimization will add many noisy words or data perturbations, so that meaningful words will be masked by tail noise. However, this situation is common in text topic analysis. 

The numerical experiments demonstrate that the convex NMF model has effectively improved the sparsity and stability of the topic clustering results in this paper, thus optimizing the topic clustering and topic ranking. The proposed two-stage dynamic topic analysis approach effectively detects emerging topics such as COVID-19 and digital twins and provides a more comprehensive analysis of clustering differences between traditional and emerging research topics. The ranking variations of emerging topics in NMF and cNMF further highlight the correlation between emerging and traditional topics.

\section{Conclusions}
The proposed two-stage convex nonnegative matrix factorization model demonstrates efficacy and stability in addressing the complexities of dynamic topic analysis in academic journals. Through the optimization of sparsity, notable improvements are achieved in topic consistency and clustering stability. Specifically, for long-term analysis, the model shows significant enhancements in topic consistency. Moreover, when considering sparsity levels of 0.4, 0.6, and 0.9, the degree of variation in topic rankings decreases by 24.51\%, 56.60\%, and 36.93\% respectively. In the dynamic analysis of 2020-2022, the model successfully uncovers emerging topics such as digital twins and the COVID-19 virus. It also enhances the coherence of topic relationships, providing a more reasonable representation of the interplay between research subjects. The extensive numerical experiments conducted validate the effectiveness and stability of the proposed method. These findings contribute to the uncovering of emerging topics and the improvement of topic relationship coherence, offering valuable insights for researchers across diverse domains.

\section{Code Availability}
The implementation of the methods presented in this paper will be available upon publication as open-source software on GitHub: \url{https://github.com/meetyangyang/CDNMF}. The code is released under the MIT License, which allows free use, modification, and distribution. Please refer to the repository for further details, including setup instructions and example usage. 

\printbibliography

\newpage  % 强制换页

\begin{IEEEbiography}[{\includegraphics[width=1in,height=1.25in, clip, keepaspectratio]{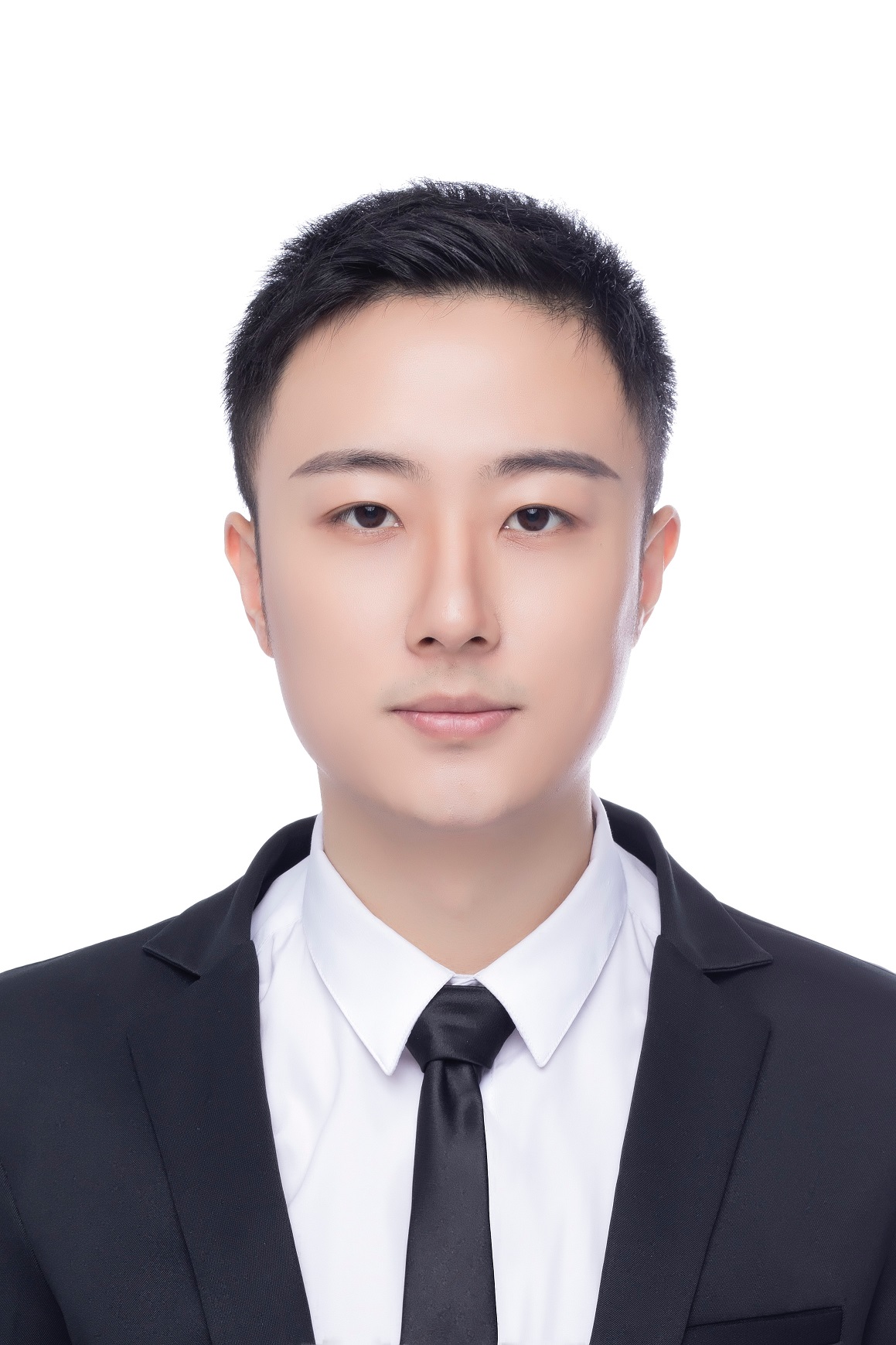}}]{Yang Yang} received the B.S. degree in Information and Computational Science, the M.S. degree in Applied Statistics, and the Ph.D. degree in Control Science and Engineering from Northeastern University, China, in 2013, 2015, and 2023, respectively.

He is currently an Associate Professor at China Criminal Police University. He was a Visiting Scholar with the Center for Advanced Process Decision-making at Carnegie Mellon University, Pittsburgh, PA, USA. He specializes in the exploration, innovation, and application of big data and artificial intelligence technologies, with a focus on advancing industry, economics, and security. His work bridges cutting-edge research with real-world solutions, addressing critical challenges in these fields. His research interests include big data analytics, data mining, multimodal models, mixture-of-expert systems, lightweight data models, and convex optimization.
\end{IEEEbiography}

\begin{IEEEbiography}[{\includegraphics[width=1in,height=1.25in, clip, keepaspectratio]{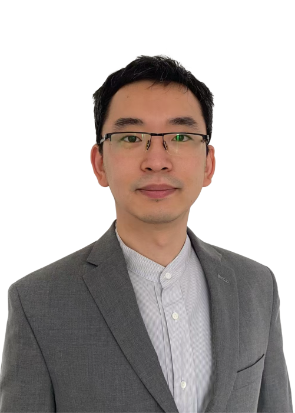}}]{Tong Zhang} received the B.S. degree in Chemical Engineering from Tianjin University, Tianjin, China, in 2013, and the M.S. and Ph.D. degrees in Chemical Engineering from Carnegie Mellon University, Pittsburgh, PA, USA, in 2014 and 2019, respectively.

Since 2022, he has been a Senior Research Engineer at the Beijing Institute for General Artificial Intelligence, where his work focuses on advancing artificial intelligence technologies. His research interests encompass large language models, text mining, machine learning, and optimization modeling.
\end{IEEEbiography}
% \vspace{-60mm} % 手动减少间距
\begin{IEEEbiography}[{\includegraphics[width=1in,height=1.25in, clip, keepaspectratio]{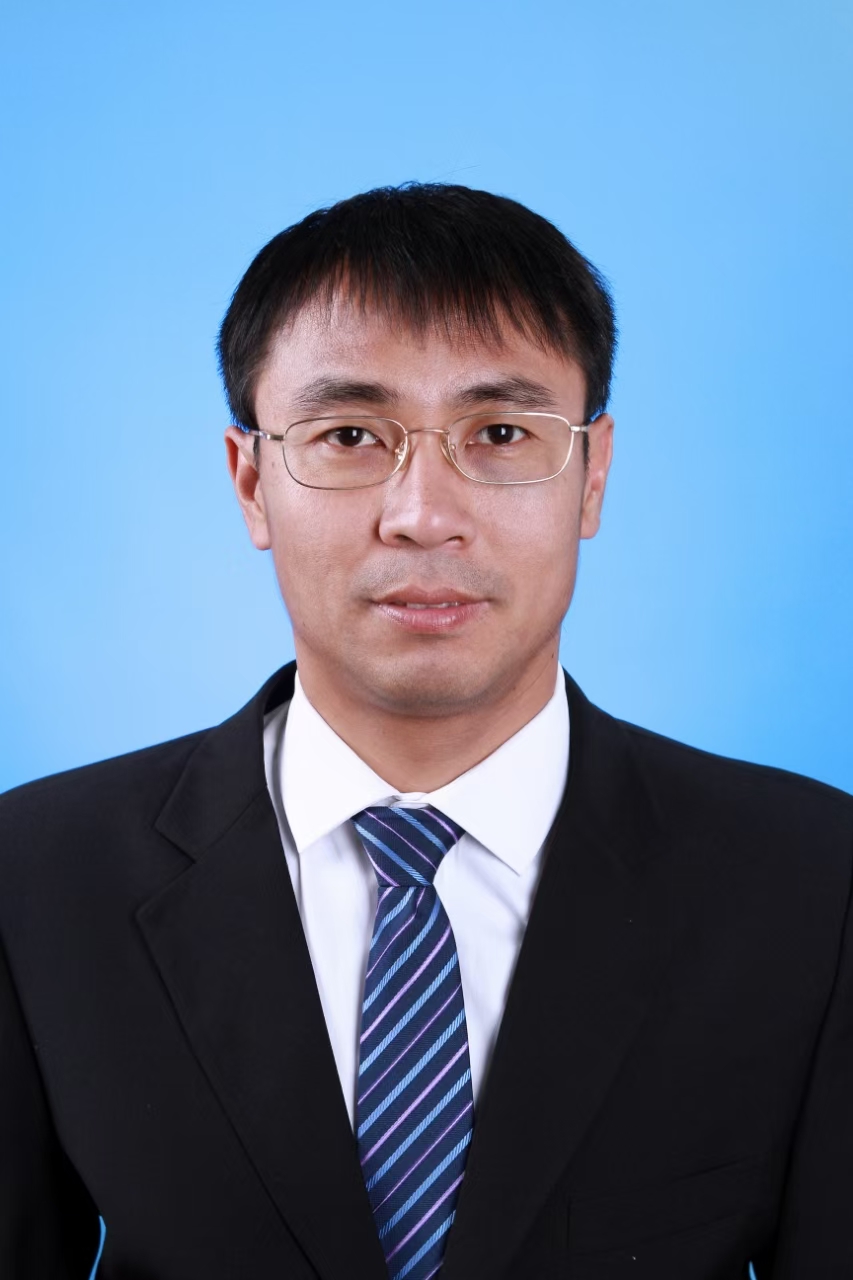}}]{Jian Wu} received the B.S. degree and M.S. degree in Applied Mathematics, with a focus on Probability Theory and Mathematical Statistics, from Northeastern University in 2002 and 2006, respectively, and the Ph.D. degree in Statistics from Beijing University of Technology in 2015.

He is currently a Lecturer at the Liaoning Engineering Laboratory of Data Analytics and Optimization for Smart Industry, Northeastern University. His research interests include non-parametric statistics, functional data analysis, industrial data modeling, statistical modeling, and convex optimization methods.
\end{IEEEbiography}
% \vspace{-60mm} % 手动减少间距
\begin{IEEEbiography}[{\includegraphics[width=1in,height=1.25in, clip, keepaspectratio]{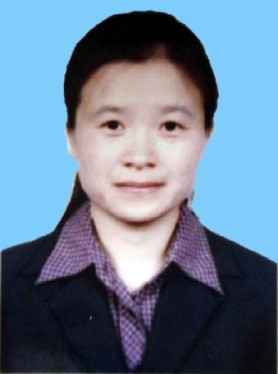}}]{Lijie Su} received the B.S. and M.S. degrees in Mechanical Engineering from Northeastern University, China, and the Ph.D. degree in Mechanical Engineering from the Shenyang Institute of Automation, CAS, in 1997, 2001, and 2005, respectively.

She is currently an Associate Professor at the Key Laboratory of Data Analytics and Optimization for Smart Industry, Northeastern University, China. Her research interests include Mathematical Modeling, Convex Optimization, MINLP, and Machine Learning. She has published over 20 papers in academic journals, including \emph{Computers and Chemical Engineering}, \emph{International Journal of Production Research}, and \emph{Computers \& Industrial Engineering}.

\end{IEEEbiography}

\end{document}